# First life in primordial-planet oceans: the biological big bang

Carl H. Gibson [1,2]

[1] University of California San Diego, La Jolla, CA 92093-0411, USA
[2] cgibson@ucsd.edu, http://sdcc3.ucsd.edu/~ir118

N. Chandra Wickramasinghe[3,4]

[3] Cardiff Centre for Astrobiology, 24 Llwynypia Road, Lisvane, Cardiff CF14 0SY
[4] NCWick@gmail.com

Rudolph E. Schild [5,6]

[5] Center for Astrophysics, 60 Garden Street, Cambridge, MA 02138, USA
[6] rschild@cfa.harvard.edu

**Abstract:** A scenario is presented for the formation of first life in the universe based on hydro-gravitational-dynamics (HGD) cosmology. From HGD, the dark matter of galaxies is H-He gas dominated planets (primordial-fog-particle PFPs) in million solar mass clumps (protoglobularstar-cluster PGCs), which formed at the plasma to gas transition temperature 3000 K. Stars result from mergers of these hot-gas-planets. Over-accretion causes stars to explode as supernovae that scatter life-chemicals (C, N, O, P, S, Ca, Fe etc.) to other planets in PGC clumps and beyond. These chemicals were first collected gravitationally by merging PFPs to form H-saturated, high-pressure, dense oceans of critical-temperature 647 K water over iron-nickel cores at ~ 2 Myr. Stardust fertilizes the formation of first life in a cosmic hot-ocean soup kitchen comprised of all planets and their moons in meteoric communication, $> 10^{100}$ kg in total. Ocean freezing at 273 K slows this biological big bang at ~ 8 Myr. HGD cosmology confirms that the evolving seeds of life are scattered on intergalactic scales by Hoyle-Wickramasinghe cometary panspermia. Thus, life flourishes on planets like Earth that might otherwise be sterile.

1. Introduction

Over many decades Hoyle and Wickramasinghe have compiled a powerful case to support the idea that life on Earth must be of extraterrestrial origin (Wickramasinghe 2005). Why should Earth be the only planet with life? Astrophysical spectra show dust clouds dominated by polycyclic-aromatic-hydrocarbons (PAH), strongly indicating that biological processes are commonplace in the Galaxy (Gibson and Wickramasinghe, 2010). The theory of cometary panspermia (Hoyle and Wickramasinghe, 2010) provides the logical mechanism for distribution of the seeds of life, but how are sufficient numbers of comets and meteors formed? How, when, and where did life begin in the first place, how widely is it distributed, and are life forms likely to be similar everywhere?

In earlier papers we have suggested an origin of life in primordial planets that began their condensations some 300,000 years after the big bang (Gibson & Wickramasinghe, 2010). This model is predicated by physical conditions prevailing in the HGD (hydro-gravitational-dynamics) cosmological model (Gibson & Schild 2009). Viscous stresses prevent gravitational structures from forming during the plasma epoch until 30,000 years (Gibson 1996), when protosuperclustervoids are triggered at the Hubble scale $ct$ and expand at sound speed $c/3^{1/2}$. Weak turbulence at void boundaries produces protogalaxies fragmenting along vortex lines just prior to the plasma to gas transition at 300,000 years. The kinematic viscosity $\nu_{gas}$ decreases by $\sim 10^{13}$ from photon-viscosity values $\nu_{plasma} \sim 4 \times 10^{25}$ m$^2$ s$^{-1}$. Gas proto-galaxies fragment into Jeans-mass clumps of Earth mass planets (Gibson 2000, 2010). Stars form by planet mergers in the clumps and explode to produce chemicals, oceans, and life (Fig. 1) in the hot-gas planets (Gibson, Wickramasinghe & Schild 2010).

In the present paper we explore aspects of this HGD model in greater detail, suggesting that the conditions within ultra-high pressure interiors of primordial planets at 647K provide optimal conditions for life's first origin. Given the large cosmological volumes available and the numerous panspermial mechanisms for communication of life templates provided by HGD cosmology, first life most plausibly appeared and was scattered among $10^{80}$ available planetary interiors produced by the big bang in the time period t = 2-8 Myr. We term this period the biological big bang.

2. **Hydrogravitational Dynamics Cosmology**

A major conclusion of hydro-gravitational-dynamics (HGD) cosmology is that viscous forces cannot be neglected in the formation of cosmic structures, Gibson (1996). During the plasma epoch following the big bang, viscosity triggers fragmentation first at 30,000 years ($10^{12}$ s) on $\sim 10^{46}$ kg mass scales of thousands of galaxies and finally at 300,000 years at galaxy mass scales $\sim 10^{43}$ kg. Protogalaxies of plasma fragment at plasma-to-gas transition at two mass-scales; the sonic Jeans-scale $\sim 10^{36}$ kg, and the viscous-gravitational Earth-scale $\sim 10^{24}$ kg. The density $\rho_0 \sim 4 \times 10^{-17}$ kg m$^{-3}$ and rate-of-strain $\gamma_0 \sim 10^{-12}$ radians s$^{-1}$ are fossil remnants of the plasma when viscous-gravitational structures first formed at $10^{12}$ s.

These trillion-planet-clump PGCs were independently identified as the dark matter of galaxies by Schild (1996) from the intrinsic twinkling frequency and brightness differences between galaxy-microlensed quasar-images adjusted for their different time delays. The difference twinkling frequency is rapid, indicating planet mass objects, not stars, dominate the mass of galaxies. The stable image brightness difference shows that the dark matter planets are in million-solar-mass clumps. Thus, cold-dark-matter cosmology fails to describe the gravitational formation of all the basic cosmological structures observed and how they evolve, Gibson (2000, 2010).

### 3. First Life in HGD Cosmology

Figure 1 outlines schematically the HGD scenario for first life that is envisaged and developed in this paper. The hot-hydrogen-planets (right) process the first chemicals supplied by supernovae.

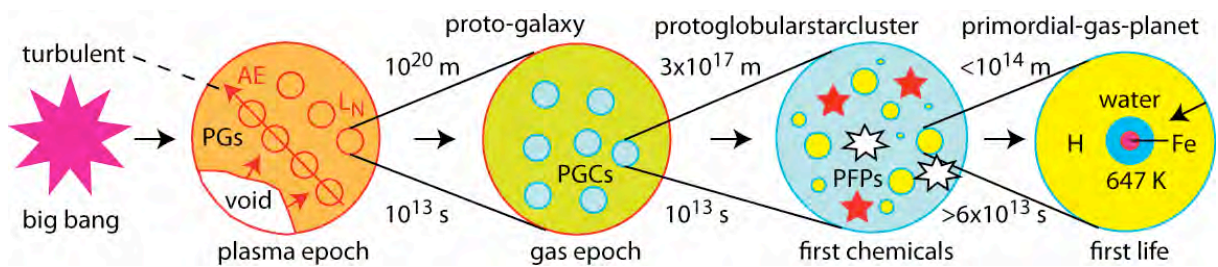

Fig. 1. Hydrogravitational dynamics HGD cosmology leads to conditions for first life. A turbulent big bang event (left) gives a superviscous plasma epoch where weakly turbulent PG protogalaxies develop at expanding superclustervoid boundaries just before the transition to gas. The spin direction (axis of evil AE) is a remnant of big bang turbulence (Schild & Gibson 2008). In the gas epoch (center) the PGs fragment into protoglobularstarcluster PGCs and PFP primordial planets. Stars form and die within the PGCs as supernovae that produce the first chemicals, including water. Critical temperature water in numerous PFP planets communicating by cometary panspermia fosters first life (right), spreading throughout the Galaxy before the oceans freeze as the biological big bang.

HGD cosmology in Fig. 1 begins (Gibson 2004, 2005, 2010) with a turbulent big bang event (red 9 point star) that leaves a spin-remnant "axis-of-evil" AE, Schild & Gibson (2008), inflation of space, and nucleosynthesis of H and He plasma, with plasma-mass about 3% of the total, 97% neutrinos, and 0% dark energy $\Lambda$, Gibson (2009ab). The smallest gravitational structures formed are protogalaxies PGs, which promptly fragment at plasma-gas transition into meta-stable PGC clumps of primordial-gas-planets PFPs. These planets condense and merge to form stars, and overfeed the stars as comets to form supernovae (white seven point stars).

The role of supernovae in contributing to dust and molecules in interstellar space was first discussed by Hoyle and Wickramasinghe (1968,1970,1988). Figure 2 shows the equilibrium gas phase abundances of the principal molecular species in a calculation appropriate to a supernova producing elements with solar relative abundances in an envelope possessing total hydrogen density $10^{18}$ m$^{-3}$.

($10^{-9}$ kg m$^{-3}$). The dashed segments for Fe, MgO, SiO$_2$ indicate that these solid phases have condensed and are in equilibrium with the gaseous component at the temperatures indicated. Note that iron condensed as a metal (whiskers) rather than its oxides according to this calculation. We note also that the gas phase molecular mixture in the ejecta includes molecules crucial in prebiotic chemistry – H$_2$O, CO$_2$, NH$_3$, H$_2$CO, HCN. The formation of SiO$_2$ (clay) particles is also predicted and may also have a relevance to prebiology.

The molecules and solid particles such as are shown in Fig. 2, formed in supernovae near the central regions of PGC clouds, would be rapidly expelled and incorporated in the proto-planet clouds (PFPs) condensing further out in each PGC system. PFP planets impregnated with molecules as well as iron particles and clays may be expected to develop layered structures due to gravitational settling on free-fall timescales $\sim(\rho G)^{-1/2}$ less than a million years. Melted by friction and gravitational potential energy of mergers, iron-nickel and molten rocky cores will be the first to form, surrounded by hot organic soup oceans under very high pressures arising from overlying massive H-He atmospheres.

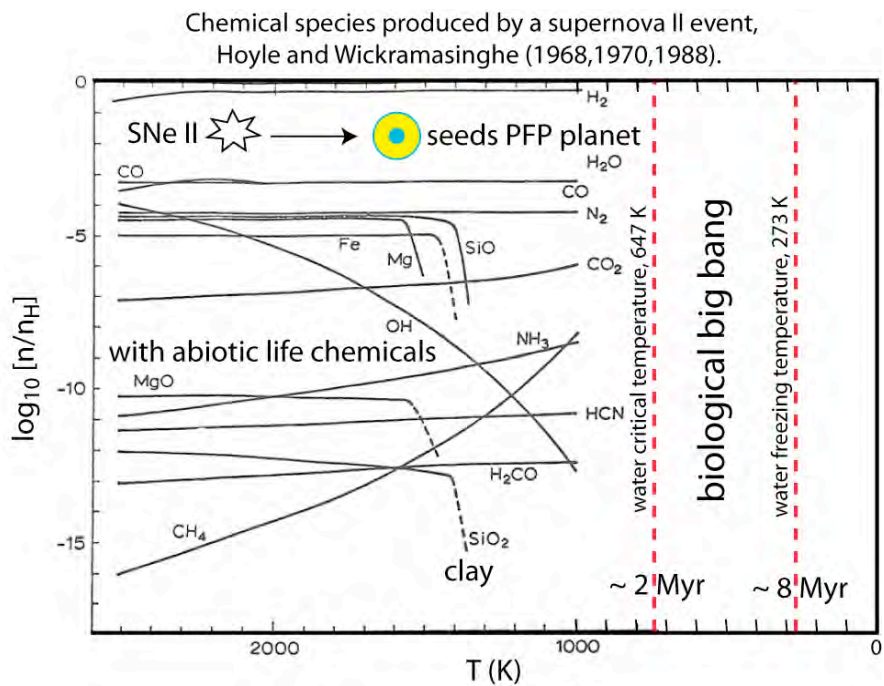

Fig. 2. Equilibrium molecular abundances in an atmosphere of density $10^{18}$ H-atoms m$^{-3}$, with solar abundances of elements. The abundances are relative to the total hydrogen density. The dashed segments for Fe, MgO, SiO$_2$ indicate that the solid phase has formed and is in equilibrium with the gaseous component at the temperatures indicated.

What makes primordial-fog-particle (PFP) planets perfect as the source of first life is their hydrogen-helium composition and the 3000 K temperature of the universe at $10^{13}$ s. Figure 3 shows the initial condition of a PFP gas planet at a time ~ 2 Myr when the universe temperature matches the critical point temperature of water at 647 K, Wickramasinghe, C., Wallis, J., Gibson, C.H. & Schild,

R. E. (2010). Water at its critical temperature is dominated by apolar linear and inverse dimers, Bassez et al. (2003) that are capable of strongly dissolving the otherwise weakly soluable apolar molecules of organic chemistry such as C, CO, $CO_2$, $CH_4$ and HCN, along with large H concentrations from the massive planet atmospheres.

Organic chemistry is complex, with ~ $2x10^7$ known organic chemicals and reaction networks depending critically on temperature as well as catalysts. The high-temperature, high-pressure, high-density reducing conditions provided by PFP planets with critical-temperature oceans, such as that shown in Fig. 3, permits maximum reaction rates after water condenses as liquid at 647 K.

With the addition of clay particles in suspension acting as catalysts and primitive templates we have chemical factories within large numbers of PFPs with the *potential* for life. The mean separation between "factories" is the nearly constant separation of PFPs within PGCs, about $10^{14}$ m (1000 AU). With some $10^{80}$ such interconnected chemical factories in communication via meteors and bolides, a cosmic primordial soup unsurpassable in scale is existent from 2 to ~8Myr after the cosmological big bang. It is in this setting that a cosmic origin of life must occur. Life patterns may be subsequently deep frozen as dormant cells in suspended animation that remain in this state until certain planets are heated to more friendly temperatures in recent cosmological epochs.

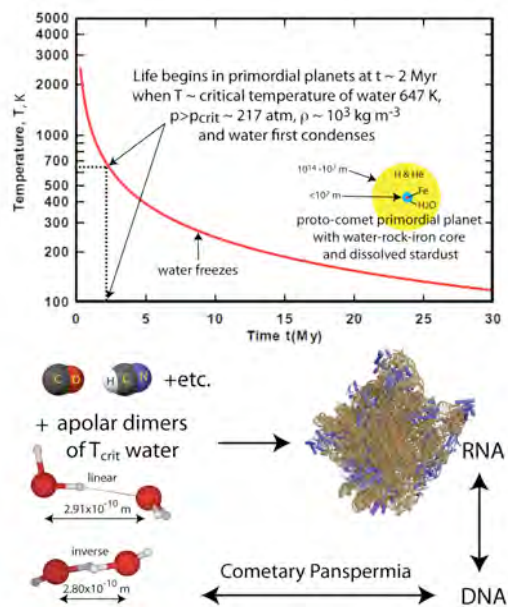

Fig. 3. Scenario for the formation of prebiotic chemicals and first life in a PFP gas planet after ~ 2 Myr. To accomplish the transition from basic organic chemicals to self-replicating molecules like RNA in the age of the universe implies an early start and numerous high temperature oceans connected on cosmic scales by cometary panspermia. PFP oceans were likely colored by the reaction of hydrogen cyanide with iron to form the dye Iron(II,III) hexacyanoferrate(II,III) called Prussian Blue.

Stars form by mergers of PFP planets within PGC clumps. A million PFP planets are needed to form a star. As planets surrounding a star slowly continue merging as planet-comets, a white-dwarf forms to continue conversion of supplied hydrogen and helium to carbon, oxygen and nitrogen until the star mass reaches the unstable Chandrasekhar limit of 1.44 solar mass. The spinning WD forms a planetary-nebula PNe of evaporated planets and finally explodes as a Supernova Ia event, spreading C, N, O, P, etc. life chemicals throughout the PGC clump and beyond. But this takes billions of years. Near proto-galaxy cores where sticky PGCs merge to high densities, very rapid planet-comet accretion rates mix away carbon-star cores to form elements up to iron-nickel, producing superstars that implode in less than a million years as Supernova II events. Superstars provide the first-life chemicals for hot-PFPs, hot-Jupiters, and other hot-fragment masses produced as a million $10^{24}$ kg primordial-gas-planets merge to make each $10^{30}$ kg star.

Stars everywhere and at all times form by mergers of PFP planets within PGC clumps, not from gas clouds as currently supposed. Superstars, with masses exceeding 50 solar, fully evolve to a type II supernova on time scales of one Myr and disperse r-process elements, rich in CNO and all metals, throughout PGCs and beyond in a short time. Thereby the early chemical enrichment of the universe evident in measured abundances of heavy elements in distant galaxies, and the chemical precursor to the emergence of life, occurred early in the universe. The prospects for wide distribution of biological materials in meteors is substantially enhanced because the million times higher density of the universe at the time favored collisional interaction. A secondary period of enrichment occurs approximately a billion years after the big bang, when an aging population of approximately solar-mass stars evolves to red giant phase and widely distributes C-N-O elements by stellar winds and supernovae to a universe at lower density. These stellar properties occur primarily within the primordial clumps of planets that gradually disperse into the dark matter halos and disks of galaxies.

Planet mergers forming stars within a galaxy constantly recycles the reaction products. The reducing environment of HGD star formation explains why solar system planets have iron cores. For example, most of planet Mercury mass is in its $>10^{23}$ kg metallic iron core. Earth's iron core is $\sim 10^{24}$ kg. Iron-nickel-metallic meteorites are ~90% of the $5 \times 10^5$ kg total on Earth. So much metallic iron in planet cores is quite inexplicable according to standard star-driven planet formation models where a handful of hydrogen free planets are formed (only 8 for the sun) rather than the HGD prediction of $3 \times 10^7$ PFP planets per star. Hundreds of known exo-planets confirm the HGD prediction.

Figure 4 illustrates HGD mechanisms in action in a dramatic way within the Helix Planetary Nebula, thus supporting cometary-panspermia as an ongoing contemporary process within the galaxy (Gibson et al. 2010).

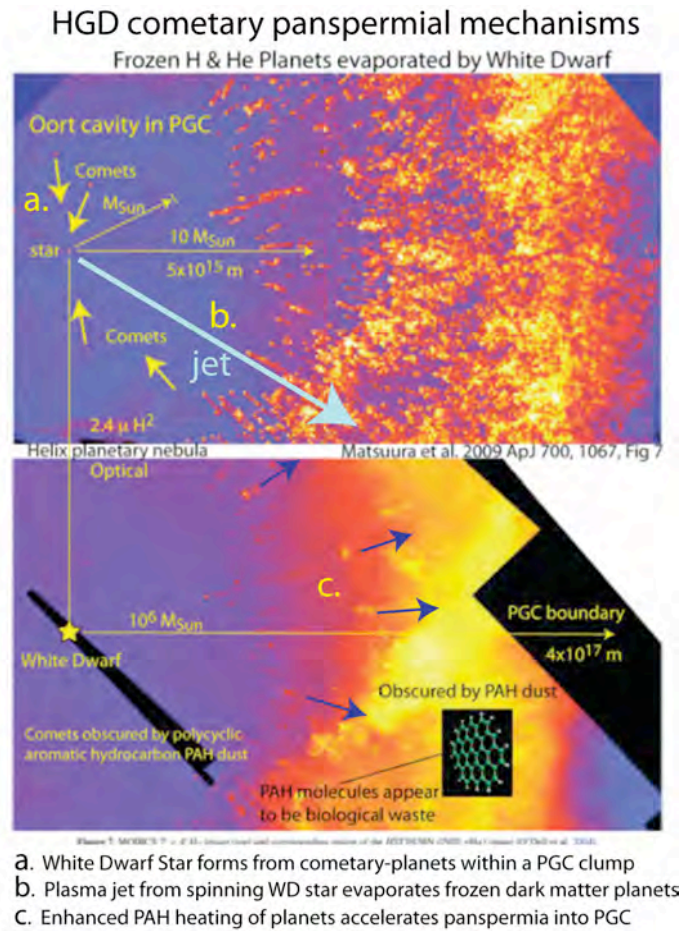

Fig. 4. Helix planetary nebula viewed at an infrared frequency sensitive to hydrogen gas 2.4 m shows ~ 40,000 partially evaporated dark matter planets surrounding its central white dwarf star (top). Comet-planets shown in 4a. have smaller atmospheres and wakes, so the plasma jet (blue arrow) is unable to prevent accretion 4b. Planets with heating enhanced by PAH 4c. grow larger atmospheres, ~ $10^{13}$ m or 100 AU, and are expelled radially by radiation pressure 4c. Hubble space telescope optical views (bottom) are obscured by PAH dust from evaporated planets. PAH dust from the planets appears to be generated biologically. If so, such radial transport of microorganisms and other life patterns is termed cometary-panspermia.

Self replicating chemical systems such as RNA and DNA in micro-organisms develop enzyme catalysts to increase their replication speeds. Seeds of life deposited on otherwise sterile planets convert nearly all available carbon to organic forms. The large number of planets per star and the powerful HGD mechanisms for wide scattering of the templates of life on cosmic scales will tend to homogenize the range of life forms in the universe. Even if PFP planets are cosmically isolated, physical and chemical constraints suggest that approximately the same life forms will develop.

## 4. Theories of first life formation

First life is assumed to have occurred on Earth since we see life everywhere and are taught to believe no convincing evidence exists for extraterrestrial life. Copernicus convinced the scientific community, and finally even the Vatican, that the Earth is not the center of the universe, and has recently been buried in hallowed ground. However, with respect to living organisms the scientific community remains in a pre-Copernican position, fiercely resisting concepts that DNA and many diseases and possibly antigens on Earth are continuously supplied by extraterrestrial sources. Because life exists profusely on Earth it is assumed that the formation of life must be a rather trivial matter, and that possibly life began spontaneously on this planet in several ways since life has existed from its beginning 4.6 Gyr ago. However, all laboratory attempts to create life have consistently failed and all serious attempts to model probabilities suggest they will always fail. Life chemistry is too complex to replicate with present technologies on human-life-time scales.

The standard Earth-based models for the origin of life, inspired by early suggestions of Haldane (1929) and Oparin (1957), are all based upon an unprovable article of faith. Faith comes in by positing the existence of complex chemical pathways that are yet to be discovered, reaction networks that are somehow capable of bridging the difficult gap between chemistry and biology. The total volume of possible Earth-based venues that serve as cradles of life, be it on the edges of primitive oceans or in deep-sea thermal vents, is minuscule compared to venues that are available in the wider space-time of the universe. Fred Hoyle in his classic book and lectures "Frontiers of Astronomy", Maddox (2003), proposed the entire solar system as a better venue than Earth for the origin of life (Hoyle, 1952), and Hoyle and one of the present authors (NCW) extended the totality of cosmic cradles to include $\sim 10^{22}$ comets in the entire galaxy (Hoyle and Wickramasinghe, 1984; 2000). Although a geocentric position is avoided in such models, the sheer scale of the improbability of an origin of life on Earth points to even wider cosmic horizons.

Serious estimates of first life probabilities give superastronically small probabilities under all conceivable Earth conditions, Wickramasinghe (2005). The continuous creation cosmology required to reconcile the observations without resorting to miracles, Hoyle et al. (2000), is rendered unnecessary by HGD cosmology, as shown in Figure 5. Cosmologies that accumulate multiverse life information in the DNA, Joseph (2010), are also unnecessary

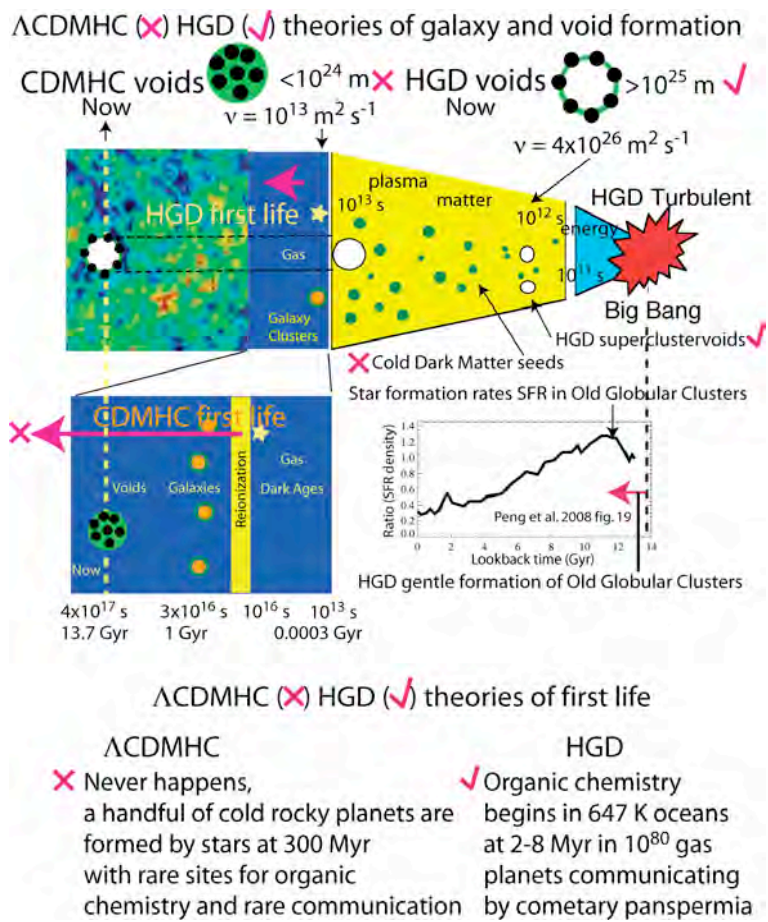

Fig. 5.  First life scenarios for hydrogravitational dynamics HGD cosmology versus standard ΛCDMHC cosmology.  At the top, void sizes and old globular star cluster evidence favor HGD cosmology.  At the bottom, the cosmologies are compared with respect to the formation of first life and eventually life as we see it on Earth.  HGD suggests life is not only possible but probable, and is spread widely in the universe by cometary panspermia as anticipated by Hoyle and Wickramasinghe.

## 5.  Click Chemistry of first life

Organic chemists faced with the problem of synthesizing useful drugs have discovered the most reliable methods follow basic principles intrinsic to biological chemistry to maximize yield and minimize time and chaos, termed "click chemistry" by K. Barry Sharpless, Kolb et al. (2001).  Figure 6 summarizes the basic ideas.

The thermodynamic basis of click chemistry reactions is their high values of thermodynamic driving force, usually greater than 20 kcal/mol.  This reduces the numbers of unwanted byproducts.  Following the example of RNA and DNA chemistry, molecular discovery processes employ enzymatic polymers as selective catalysts to speed the synthesis and avoid collapse of the reactions into chaos.

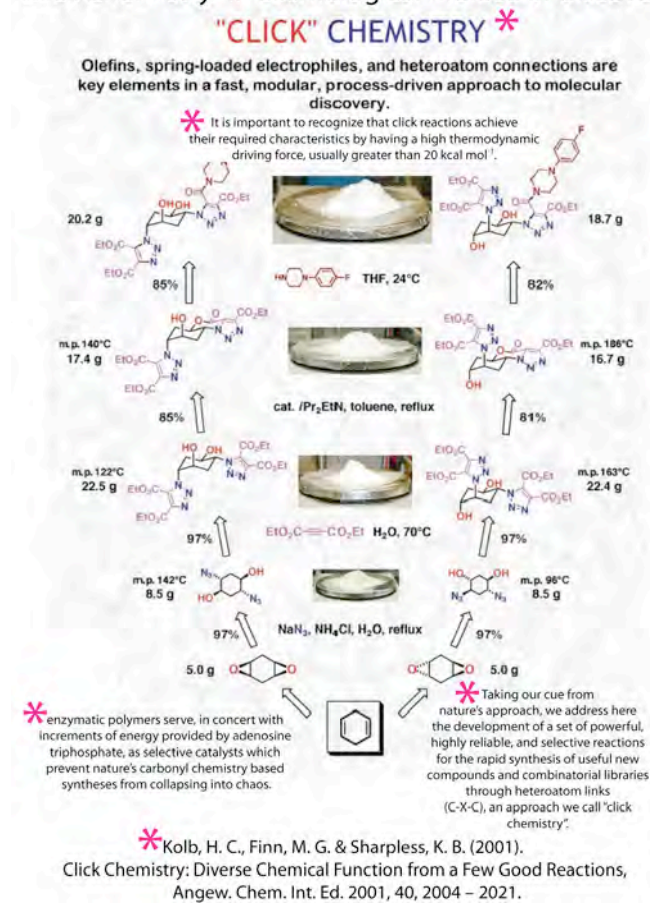

Fig. 6. Efficient chemical synthesis of molecular compounds in the laboratory require "click chemistry" methods. Cosmological molecular discovery in the formation of first life is constrained by similar, if not identical, principles. For this reason, it seems likely that independent cosmological systems may produce virtually identical life forms.

Considering the high speed of high temperature reactions expected in HGD hot gas planets and the large numbers of such nearly identical planets and their clumps in cosmic communication, it seems reasonable to expect from click chemistry similar outcomes in their productions of living organisms, starting at ~2 million years as oceans first condensed at the critical water temperature, and slowing down as oceans freeze at the surface ~ 8 million years after the big bang.

### 6. Evidence of extra-terrestrial life

Carbonaceous meteorites Orgueil and Murchison provide very strong evidence for extraterrestrial life. Figure 7 shows microscopic images of cyanobacterial filaments presented by NASA astrobiologist Richard B. Hoover at the 2010 SPIE astrobiology conference in San Diego. Hoover concludes "that the filaments found in these meteorites provide clear and convincing proof for the existence of extraterrestrial life and support the hypothesis of an exogenous rather than an endogenous origin of Earth Life" in his website given in Fig.7 (bottom). Both meteorites are older than Earth and suggest a more sophisticated organic biosphere has evolved on the planets that produced them.

Evidence includes the amino acids, nucleotides, and other life-critical biomolecules found in both carbonaceous meteorites. An excess of L-amino acids, a property of the proteins in all living organisms, is consistent with life, and has no known explanation by abiotic production processes (which yield equal numbers of the D- and L- forms). The complexity of organic chemicals (millions) present in Muchison far exceeds terrestrial levels, Schmitt-Kopplin, P. et al. (2010).

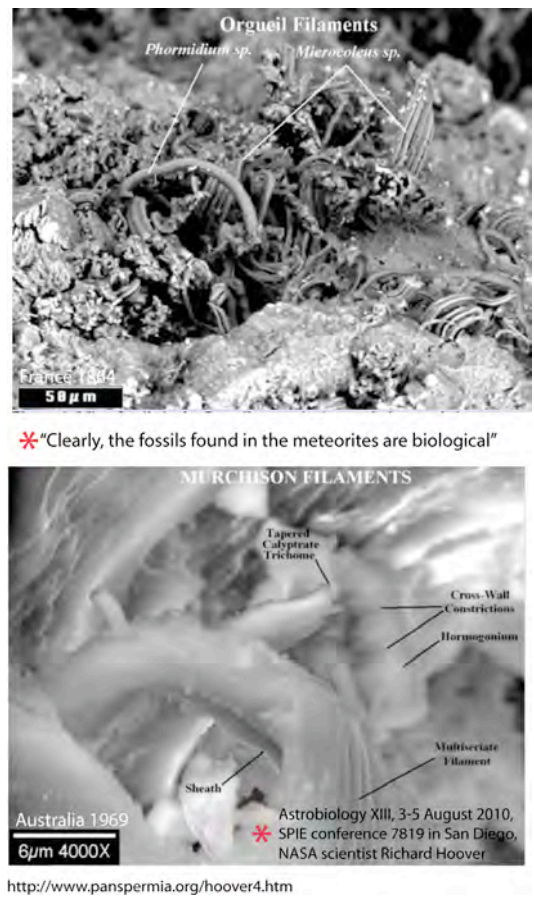

Fig. 7. Cyanobacteria fossils observed by Richard Hoover in Orgueil and Murchison meteorites.

Evidence of clearly extraterrestrial life with possibly reversed chirality DNA was presented at the 2010 San Diego astrobiology conference, Figure 8. A meteor passed close over the South Indian state of Kerala in 2001 during the monsoon season, producing a sonic boom and triggering scattered outbursts of rain with distinctive red coloration, Louis & Kumar (2003). Some $5 \times 10^5$ kg of meteoric material is estimated to have been deposited in the atmosphere. The red rain organism turns out to by an amazing hyperthermophile that survives temperatures in excess of the critical temperature of water 647 K, and displays a maximum growth rate at 573 K (300 °C), Fig. 8 (top).

Examples of red rain over history are rather common, McCafferty, P. (2008), but no previous attempt has been made to identify why the rain is red. According to the Louis studies the red mode of the organism is a resting phase. The daughter organisms are yellow or colorless, but grow and clump to form red spores, Fig. 8 (bottom right), similar to those originally collected (photo).

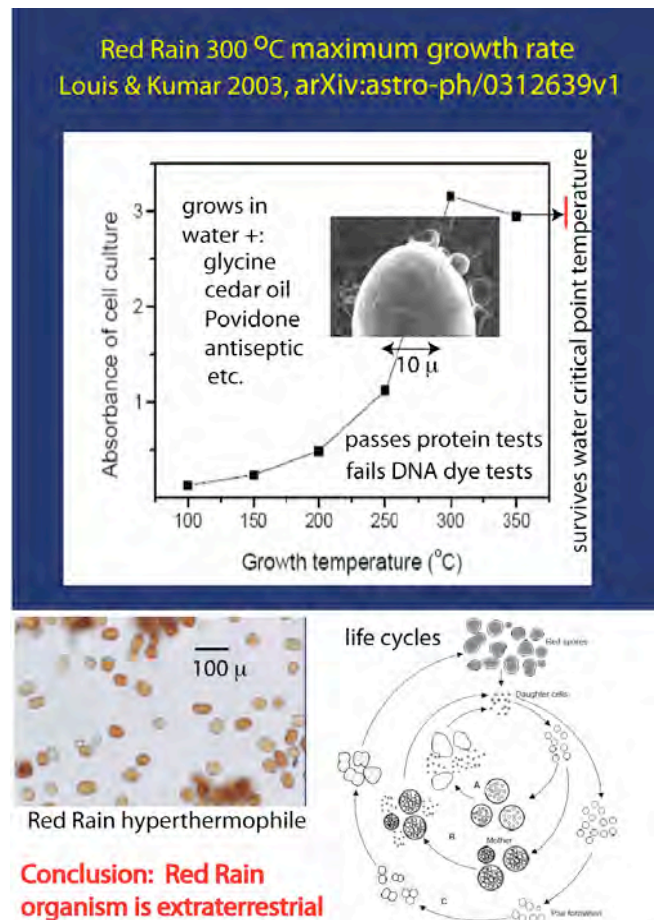

Fig. 8. Tests of Red Rain hyperextremophile organism, Louis & Kumar (2003), show survival and maximum growth rates at temperatures exceeding the water critical temperature (top). Life cycles (bottom right) show daughter cells emerging from mothers, thick protective walls (transmission electron microscope), and forming cysts, but no DNA is detected by standard dye tests. A scanning electron microscope image (top insert) and growth tests to 121 $^{o}$C support the L&K Conclusion that the Red Rain organism is extraterrestrial.

Follow up studies produced the scanning electron microscope image of the organism at the top of Fig. 8, Gangappa, R, Wickramasinghe, N. C., Kumar, S. & Louis, G. (2010). Transmission electron microscope studies show thick cell walls, and support many aspects of the life cycles shown in Fig. 8 (bottom right) involving daughter cells emerging from the interior of larger mother cells. Protein tests confirm the organism is biological, but standard DNA dye tests fail. DNA dye tests involve chiral molecules and would fail to indicate DNA if the red rain organism is an extraterrestrial representative of the shadow biosphere, where DNA chirality is reversed. It is a challenge to astrobiologists to develop shadow-biosphere DNA dye tests.

### 7. Discussion

HGD cosmology provides strong support to the Hoyle-Wickramasinge hypothesis of cometary panspermia within known laws of physics and fluid mechanics. The standard $\Lambda$CDMHC cosmology is incompatible with life developing anywhere in the universe, forcing the assumption of mira-cles to explain life on Earth or the need for new physics or multiverse assumptions. Estimates are made using HGD mechanisms of how and when first life appeared that make miracle and multiverse models unnecessary.

As shown in Figs. 1, 2 and 3, the most likely time for life forms to first have a chance to appear is at 2-8 million years after the big bang when the first stars and supernovae of HGD cosmology have begun to spread life chemicals, particularly water, to HGD Jeans-mass clumps of primordial gas planets, and the water is able to condense at its critical temperature of 647 K and has not yet begun to freeze at 273 K. Conditions to form and distribute life are optimum, producing a very rapid period of organic chemistry activity and exchange we term the "biological big bang" event.

Previous to 2 million years, such hot planets would simulate steam hydrocarbon reforming units that operate at 1000-1300 K at high pressures with nickel catalysts to reduce petrochemicals to hydrogen and carbon monoxide. Soon after 2 million years and the condensation of oceans, complex life processes should develop and be widespread throughout the cosmos by cometary panspermia mechanisms illustrated in Fig. 4abc.

The length of the HGD timeline for first life shown in Fig. 5 is not certain, but should be more than a few million and less than a few hundred million to spread throughout a galaxy. The rate of development of life is accelerated by the thermodynamic principles of "click chemistry", Fig. 6. Since primordial planets in galaxy dark matter clumps develop in very uniform conditions from HGD, guided by the same laws of physics, thermodynamics and reaction kinetics, it seems unlikely that life forms developing in one region of the cosmos will be very different from life forms developed in another. Since our horizon covers only about $10^{-40}$ of the big bang cosmos, we are unlikely to be able to test this hypothesis.

The critical length scale for different life form developments is the scale of protoglobularstarcluster PGCs, which is $3 \times 10^{17}$ m from HGD. Available extraterrestrial life forms shown in Figs. 7 and 8 appear to be from two different biospheres, with DNA handedness (chirality) reversed.

## 8. Conclusions

The optimum conditions for first life to develop and be widely dispersed on cosmic scales occurs as a biological big bang event ~2-8 million years after the cosmological big bang according to HGC cosmology. This is the time period while oceans of water existed in hydrogen-helium gas planets at the critical temperature of first condensation, while maximum solubility and rapid reactions in many communicating planet oceans can foster the hydrocarbon chemistry needed for development of life as we know it, and before ocean freezing begins. Because life is prolific on Earth and would be impossible from ΛCDMHC cosmology, it is clear ΛCDMHC cosmology fails in regard to explaining life. HGD cosmology is strongly supported by the existence of life on Earth, just as HGD cosmology strongly supports the Hoyle-Wickramasinghe concept of cometary panspermia. A cosmological primordial soup cooked up in $10^{80}$ hot primordial gas planets in clumps provided by HGD easily explains why all galaxies leave polycyclic-aromatic-hydrocarbon dust trails when they are observed to pass through each other's dark matter halos (Gibson & Schild 2002, 2007a). Stars are triggered into formation by primordial planets in dark matter clumps, expelling large quantities of biological waste products from merging planets. Examples of this process in our own Galaxy are seen in the Helix Planetary Nebula (and others), Figure 4 (Gibson & Schild 2007b).